\documentclass[fleqn,usenatbib]{mnras}

\usepackage{newtxtext,newtxmath}

\usepackage[T1]{fontenc}

\DeclareRobustCommand{\VAN}[3]{#2}
\let\VANthebibliography\thebibliography
\def\thebibliography{\DeclareRobustCommand{\VAN}[3]{##3}\VANthebibliography}

\usepackage{graphicx}	
\usepackage{amsmath}	

\usepackage{listings}
\usepackage{multicol}
\usepackage{graphicx}
\usepackage{subcaption}
\usepackage{stackengine}

\title[Occultation Portal]{Occultation Portal: a web-based platform for data collection and analysis of stellar occultations}

\author[Y. Kilic et al.]{Y. Kilic$,^{1,2}$\thanks{E-mail: \href{mailto:yucelkilic1@gmail.com}{yucelkilic1@gmail.com} (YK)}
F. Braga-Ribas$,^{3,4,5}$
M. Kaplan$,^{1}$
O. Erece$,^{1,2}$
D. Souami$,^{6,7,8}$
M. Dindar$,^{2}$
\newauthor
J. Desmars$,^{9,10}$
B. Sicardy$,^{7}$
B. E. Morgado$,^{4,5,7,11}$
M. N. Shameoni$,^{12}$
F. L. Rommel$,^{4,5}$
\newauthor
and A. R. Gomes-Júnior$^{5,13}$
\\
$^{1}$Department of Space Sciences and Technologies, Akdeniz University, Campus, Antalya, 07058, Turkey\\
$^{2}$T{\"{U}}B{\.{I}}TAK National Observatory, Akdeniz University Campus, Antalya, 07058, Turkey\\
$^{3}$Federal University of Technology, Parana (UTFPR-Curitiba), Curitiba/PR, Brazil\\
$^{4}$Observatório Nacional/MCTIC, R. General José Cristino 77, Rio de Janeiro, RJ 20.921-400, Brazil\\
$^{5}$Laboratório Interinstitucional de e-Astronomia - LIneA and INCT do e-Universo, Rua Gal. José Cristino 77, Rio de Janeiro, RJ 20921-400, Brazil\\
$^{6}$Université Côte d’Azur, Observatoire de la Côte d’Azur, CNRS, Laboratoire Lagrange, Bd de l’Observatoire, CS 34229, 06304 Nice Cedex 4, France\\
$^{7}$LESIA, Observatoire de Paris, Université PSL, CNRS, Sorbonne Université, Université de Paris, 5 place Jules Janssen, 92195 Meudon, France\\
$^{8}$naXys, University of Namur, 8 Rempart de la Vierge, Namur, B-5000, Belgium\\
$^{9}$Institut Polytechnique des Sciences Avanc\'ees IPSA, 63 boulevard de Brandebourg, F-94200 Ivry-sur-Seine, France\\
$^{10}$Institut de Mécanique C\'eleste et de Calcul des \'Eph\'em\'erides, IMCCE, Observatoire de Paris, PSL Research University, CNRS,Sorbonne Universités,\\ UPMC Univ Paris 06, Univ. Lille, 77 Av. Denfert-Rochereau, F-75014 Paris, France\\
$^{11}$Universidade Federal do Rio de Janeiro - Observatório do Valongo, Ladeira Pedro Antônio 43, CEP 20.080-090 Rio de Janeiro - RJ, Brazil\\
$^{12}$Atat{\"{u}}rk University Application and Research Center for Astrophysics (ATASAM)\unskip, 25240\unskip, Erzurum\unskip, Turkey\\
$^{13}$UNESP, São Paulo State University, Grupo de Dinâmica Orbital e Planetologia, CEP 12516-410, Guaratinguetá, SP 12516-410, Brazil
}

\date{Accepted XXX. Received YYY; in original form ZZZ}

\pubyear{2022}

\begin{document}
\label{firstpage}
\pagerange{\pageref{firstpage}--\pageref{lastpage}}
\maketitle

\begin{abstract}
Recording a stellar occultation is one powerful method that gives direct information about the physical properties of the occulting solar system object. In order to obtain reliable and accurate results, simultaneous observations from different locations across-track of the projected path are of great importance. However, organising all the observing stations, aggregating, and analysing the data is time-consuming and not that easy. 
We have developed a web portal named {\it Occultation Portal (OP)} to manage all those occultation observation campaigns from a central server. With this portal, the instrumental and observational information of all observers participating in a stellar occultation campaign and the concerned data are archived systematically in a standard format. The researchers can then visualise the archived data on an event basis. The investigators can also extract the light curve for each data-set with the added reduction pipeline to the portal base. This paper describes in detail the portal structure and the developed features.
\end{abstract}




\section{Introduction}\label{sec:intro}

A stellar occultation is a celestial event that occurs when a Solar System object obscures a source from the perspective of an observer. Apart from the occultation event caused by the main body, secondary events may also be observed caused by rings, dust, or satellite in the vicinity of the main body  \citep{sicardy2011,ortiz2012,fribas2014,ortiz2017}. Similarly a gradual dis/re-appearance of the main body can be recorded revealing an atmosphere.  The duration of these occultation varies according to the size, shape, and sky motion rate of the object. When these events are recorded simultaneously from more than two locations across the occultation path, a multi-chord stellar occultation, we are able to obtain detailed information about the foreground object \citep{fribas2013,damya2020}. The multi-chord stellar occultation technique is, at present, the only method able to determine the size and shape of the small objects of the outer solar system with sub-km accuracy, exception made for a visiting spacecraft. In addition to estimating the physical properties of the foreground object, the stellar angular diameter of the occulted star may also be calculated from the occultation event \citep{wideman2009}, or an unresolved stellar companion may be discovered \citep{berard2017}. 

Before the {\it HIPPARCOS} mission \citep{hipparcos1997}, stellar occultation predictions and observations were highly challenging because of the significant uncertainties in the position of occulted stars \citep{dunham2002}. Afterwards, precise occultation predictions increased significantly thanks to the highly accurate astrometric positions provided by {\it Gaia} DR1, DR2, and eDR3 catalogues \citep{gaia2016,gaia2018,gaia2021}.

Considering the more than 15000 observations from 4400 events provided by 3300 individuals from all over the world over the last 40 years \citep{herald2020}, we are seeing an extraordinary increase of successful occultation observations that will stand for the next decades\footnote{\href{http://occultations.ct.utfpr.edu.br/results}{Database on Stellar Occultations by Small Outer Solar System Objects}} \citep{fribas2019}. Not only the {\it Gaia} stellar catalogue, but also the accurate positions of small solar system objects within {\it Gaia} next releases will strikingly increase the number of precise predictions of occultations \citep{ferreira2022}. Besides, the involvement of the Vera Rubin Observatory with the {\it LSST}\footnote{\href{https://www.lsst.org/}{Legacy Survey of Space and Time}} will also have a significant impact \citep{ivezic2019}. As a result, a remarkable amount of data will be produced. 

Another factor in increasing the number of precise and successful occultation events is the development of accurate numerical methods for their prediction such as NIMA\footnote{Numerical Integration of the Motion of an Asteroid} \citep{desmars2015} for orbit determinations of moving objects. In addition, more affordable observational equipment, increased camera sensitivity, developments in the internet infrastructure and investments in cloud services \citep{caroline2019,canalys2021} have had an increasing effect on the data size produced along with the broadband data transmission. As a result of all these developments, the contribution of amateur astronomers in the observation campaigns increased tremendously \citep{herald2020}, and most of their occultation light curves, produced from events involving the Moon and asteroids, are archived in the Occultation light curves catalogue\footnote{\url{https://cdsarc.cds.unistra.fr/viz-bin/cat/B/occ}} \citep{herald2016}.

Building on these developments, research groups such as the ERC Lucky Star project\footnote{\url{https://lesia.obspm.fr/lucky-star/}} in France, Spain and Brazil and RECON\footnote{\url{http://tnorecon.net/}} in the USA, have seen the number of successful occultations increase which has lead them to groundbreaking discoveries such as those of the rings around the Centaur (10\,199) Chariklo \citep{fribas2014} and the dwarf-planet (136\,108) Haumea \citep{ortiz2017}, as well as the invaluable support to space missions such as New Horizons as the 2017 occultation by the TNO (486\,958) Arrokoth \citep{buie2020}. In addition, the ACROSS project\footnote{\url{https://lagrange.oca.eu/fr/home-across}}, in collaboration with amateur observers, recorded the potentially hazardous asteroid (99942) Apophis using the stellar occultation technique for the first time \citep{apophis2021}. These teams are seeking support from several groups of observers such as the International Occultation Timing Association (IOTA)\footnote{\url{https://occultations.org/}}, the IOTA-European Section (IOTA-ES)\footnote{\url{https://www.iota-es.de/}} as well as other groups in Australia, New Zealand\footnote{\href{https://www.occultations.org.nz/}{The Trans Tasman Occultation Alliance}}, Japan\footnote{Japanese Occultation Information Network}, South and Central America\footnote{\href{https://ocultacionesliada.wordpress.com/}{Liga Iberoamericana de Astronomía – LIADA}}, North and South Africa, India and more\footnote{\href{https://www.euraster.net/links.html}{euraster.net - Asteroidal Occultation Sites}}. Once confirmed, the predictions and calls for observers are made public through the above cited project's websites (such as the \url{https://lesia.obspm.fr/lucky-star/}) or using tools such as the \emph{Occult}\footnote{\url{http://www.lunar-occultations.com/iota/occult4.htm}}, \emph{OccultWatcher} (OW) \footnote{\url{https://www.occultwatcher.net/}} \citep{pavlov2018_1,pavlov2018_2} and it's online version the \emph{OccultWatcher Cloud} (OWC) \footnote{\url{https://cloud.occultwatcher.net/}} feeds.

We can illustrate the importance of coordination and developments in technology through the stellar occultation campaign of the 2002 MS$_4$ object dated 8 August 2020\footnote{\url{https://lesia.obspm.fr/lucky-star/occ.php?p=39076}} which  involved 116 telescopes across Europe, North Africa, and Western Asia, with the most extensive participation ever on an event by a TNO \citep{flavia2021}. Each stellar occultation campaign nowadays requires extraordinary coordination with observers from all over the world. A systematic follow-up of such participation is essential to ensure the successful and efficient interpretation of the data, and ensuring the best scientific outcome while ensuring the participation of each observer is gratified.

Collecting observational data and details on the instrumental setup from each observer through e-mail causes time delays, useless redundancies and data loss.
Professional observatories generally produce data in FITS format, while most amateur observers prefer the video format. Generally, the researcher contacts all the campaign participants by e-mail and receives all observational information and data from the observers via various cloud services. 
Since the requested observation time-span can sometime be relatively long (about an hour, or so), the data size produced by observers may exceed the quotas of free data sharing cloud services. Moreover, the researcher collecting the data might face the problem of archiving the data coming from tens or even hundreds of observers.

In order to solve all these problems and to guarantee the participants being in the scientific publications in a systematic, practical, and cost-effective way, it became crucial to develop a common online platform where the needed data is collected and analysed (producing light curves/ photometry). In the light of this, we developed a platform called Occultation Portal (OP)\footnote{\url{https://occultation.tug.tubitak.gov.tr/}} to manage ERC Lucky Star Project observation campaigns effectively and efficiently. This paper is organised as follows. In Section \ref{sec:architecture}, we introduce the current architecture of the OP and our approaches to creating its components. In Section \ref{sec:results}, the use of the OP in current campaigns is discussed with general statistics, as well as the photometric reduction and analysis. Finally, we  discuss the future improvements, plans, and challenges for the platform in Section \ref{sec:conclusions}.

\section{The architecture \& algorithm}\label{sec:architecture}

\begin{figure*}
\centering
\includegraphics[width=\textwidth]{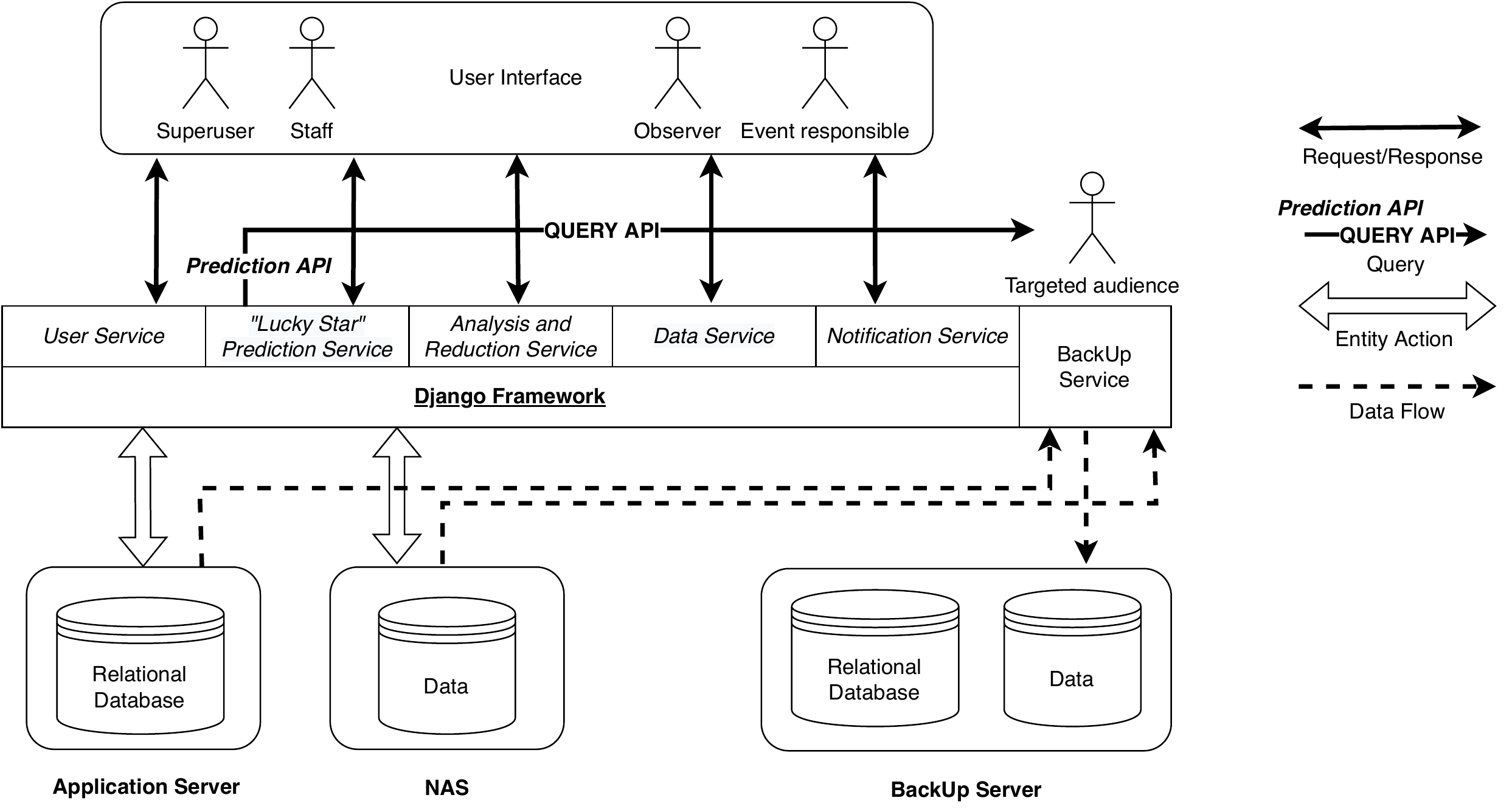}
\caption{The Occultation Portal's software architecture. Six different services running and their relationship with the environment are represented. The two-way continuous arrows show the Request/Response relationship between the services and the users via the web interface. The one-way continuous arrow represents the query from the Prediction API (Application Programming Interface) service to Targeted audience with API key. The one-way dashed arrows indicate the data flow direction of the database and observational data run by the \textit{BackUp Service}. The hollow two-way arrows represent the database and observation data \textit{Entity Actions} coming through the framework.}
\label{fig:occultation_platform_design}
\end{figure*}

The Occultation Portal (OP) is a data archiving, analysis, and visualisation portal created to systematically assist in conducting observational campaigns of the ERC Lucky Star Project as its primary goal. The entire web-based portal consists of \textit{User Service}, \textit{Prediction Service}, \textit{Analysis and Reduction Service}, \textit{Data Service}, \textit{Notification Service} and \textit{Backup Service} as shown in Figure \ref{fig:occultation_platform_design}. Through these services, and depending on the mode they are using, users can perform operations such as: adding predictions, instruments, and coordinates of observing stations, uploading data associated to any listed campaign, managing campaigns, requesting information about specific events or the use of the portal. Finally, the users can use the incorporated tools for data analysis and visualisation depending on their access levels via the web interface. To create all this infrastructure and superstructure, we preferred the Django framework\footnote{\url{https://www.djangoproject.com/}} as it enables the rapid development of the OP's back-end and front-end and works in perfect harmony with the Python programming language.

\subsection{User Service}\label{sec:user_service}

In order to ensure the security of the system and the data uploaded on the OP, users are authorised to access the OP services in four different user levels: \textit{superuser}, \textit{staff}, \textit{observer} (standard users) and \textit{event responsible}. The first three levels were added with the help of the built-in Django authentication system, while the fourth level which is more privileged than \textit{observer} user was also added to allow the user to manage event-specific campaigns. Table \ref{tab:access-levels} shows each user's access levels and authorisation groups on the OP.

\begin{table*}
\caption{Authorisations and access levels of users groups to the OP's services. The primary services available on the OP are listed as Feature. {\it Yes} indicates that a user group has access and management authority to the relevant service. {\it No} means the opposite. {\it Read-only} refers to a user group that can access a service in a read-only mode without administrative authority. Superuser(s) have administrative authority on all services. Staff users are mostly Lucky Star researchers with moderately limited controls designated by a Superuser. Observers are the OP's primary group and only authorised to register an event to provide data and information. Event Responsible(s) are the researchers who can manage all processes of a peculiar event assigned by a Superuser or Staff.}
\label{tab:access-levels}
\centering 
\begin{tabular}{l c c c c c }
\hline\hline 
\textbf{Feature}    & \textbf{Superuser} & \textbf{Staff} & \textbf{Observer} & \textbf{Event Responsible} \\ \hline
\hline 
User Service   & Yes                & No             & No                & No                         \\
Prediction Services  & Yes                & Yes            & Read-only                & No                         \\
Analysis \& Reduction Service       & Yes                & Yes            & No                & Yes              \\
Data Service        & Yes                & Yes            & No                & Yes              \\
Notification Service        & Yes                & Yes            & No                & No              \\
Backup Service (*)        & No                & No            & No                & No              \\
\hline 
(*) Service is controlled only by the server system admin.\\
\hline 
\end{tabular}
\end{table*}

Switching between user groups can only be made by superusers. Staff can also assign an event responsible for a specific prediction.

\subsection{Lucky Star Prediction Service}\label{sec:predictions}

The prediction sources that feed the Occultation Portal consist of TNOs, Centaurs, and asteroids (mostly Trojans, see Fig. \ref{fig:campaign_dynamic_classes}) selected by the researchers generally from the Rio, Meudon, and Granada groups within the framework of the Lucky Star Project. Predictions are based on the studies of \cite{assafin2010} for Pluto and its system, \cite{assafin2012} for TNOs, \cite{camargo2014} for TNOs and Centaurs. The ephemerides of the selected objects are calculated using the NIMA software, as detailed in \cite{desmars2015}, and the ephemeris database is regularly updated. Observational data used in the ephemeris calculations are currently provided from the Minor Planet Center, and Lucky Star Project observations at ESO, Pic du Midi, Calar Alto, Sierra Nevada, and Observat\'orio do Pico dos Dias. Additionally, astrometric data of the Dark Energy Survey \citep{banda2019} are used. Ephemerides of the satellites of the TNOs are taken from the Genoid project \citep{vachier2012}. The astrometric positions and proper motions of the stars in the Gaia catalogue releases are used in predictions \citep{gaia2021}\footnote{Most recently Gaia eDR3 }\footnote{\url{https://lesia.obspm.fr/lucky-star/predictions.php}}. 

The entire prediction database of the Lucky Star Project is uploaded to the Occultation Portal. Each prediction in this database carries an Occultation ID ($occ\_id$). Suppose this ID is activated via the Prediction tab in OP by authorised users, then a campaign call is made for this prediction which makes it selectable by any users from the calendar. After this stage, for this announced prediction, the OP admins can appoint a user amongst registered users in the Portal as "event responsible". The selected "event responsible person" now accesses all information about this event and manages the data on the OP.

\subsection{Lucky Star Prediction API}\label{sec:prediction_api}

Lucky Star Project presents its predictions through a website. Most professional and amateur observers can search for the prediction they are interested in here. However, this web interface is not practical for most programmers or researchers to quickly access more detailed information about the predictions. Since it is not as practical as Occultation Portal, Occult Watcher, OW Cloud, etc., an Application Programming Interface (API) service has been deployed using Django Ninja\footnote{\url{https://django-ninja.rest-framework.com/}} to facilitate access to predictions\footnote{\url{https://occultation.tug.tubitak.gov.tr/api/docs}}. A user with an API key can scan the entire Lucky Star Prediction database through the Occultation Portal for desired location, specific date and brightness range, as seen in the code listing \ref{lst:predapi}.

\begin{lstlisting}[label=lst:predapi,language=bash,breaklines=true,caption={Lucky Star Prediction API usage via curl.}]
 curl -k -X GET "https://occultation.tug.tubitak.gov.tr/api/predictions?start_date=2022-02-01&end_date=2022-02-15&mag_min=10&mag_max=18&object_name=Hektor&page=1" -H 'X-API-Key: SECRET_API_KEY'
\end{lstlisting}

Curl is preferred in the example, a terminal client used for data transfer. In the example, the Prediction API query of a user with an API KEY is sampled. Here, if desired, the start date of the prediction (\textit{start\_date}), the end date of the prediction (\textit{end\_date}), minimum magnitude value of the selected object (\textit{mag\_min}), the maximum magnitude value of the desired object (\textit{mag\_max}), solar system object name (\textit{object\_name}), requesting page number of result (\textit{page}) arguments can be changed by the {\it targeted audience} as shown in the Fig. \ref{fig:occultation_platform_design}.

\subsection{Data service}\label{sec:data_service}

The observational data is provided to the OP by observers who participated in the observational campaigns. In order to participate in a campaign, the user should register specifications of the telescope, camera, time recording system, filter, and location. The OP stores data systematically and delivers the data to the event responsible(s) and admins in a well-coordinated way using those registered information. Thus, authorised users may track and manage all this information in a table and also on OpenStreetMap\footnote{\url{https://www.openstreetmap.org}}. 

It should be noted that one of the most critical information requested on the OP is the observers' accurate location which is crucial for dependable analysis. To avoid any mistake in location information, the OP has all IAU (The International Astronomical Union) registered observatories as selectable. Furthermore, the user may select a location easily with the help of the map assistant. Astropy \citep{astropy2013,astropy2018} and geopy\footnote{\url{https://github.com/geopy/geopy}} are used together in the background of all these processes. After filling all the information in, the user is directed to the data upload section to upload the data in one of the available formats such as $.fit(s)$, $.zip$, $.rar$, $.gz$, $.7z$, $.avi$, $.adv$, $.aav$, $.ser$.

All the data uploaded on the OP is only available to researchers or groups approved by the Lucky Star Project members and collaborators on an event basis. As shown in the appendix \ref{apd:op_chords_report_page}, an observer can view other participants, excluding the uploaded data on the event attended and only view and access their own data unless the observer has Data Service privileges given in Table \ref{tab:access-levels}. The authorised user who wants to access event-based data may request the event data anytime by pressing the \textit{Request Event Data} button (see appendix \ref{apd:reported_observation}). OP queries all relevant data in the database, creates a ZIP file, sends an e-mail notification to the requesting user and system admin when the process is completed to ensure data privacy. The ZIP file contains all the event data, additional files (if provided), the ROI (Report Object Instance) file (if provided), and the auto-generated Asteroidal Occultation Report\footnote{\url{https://www.occultations.org.nz/planet/AsteroidReportFormTextV2.txt}} by the OP. This file is served by $nginx$\footnote{\url{https://www.nginx.com/}} on the server.

\subsection{Data analysis and reduction service}\label{sec:data_analysis}

All types of observational data uploaded to the Occultation Portal are stored systematically on the server in an event and user basis. In addition, each recording path is also kept in the database for fast accessing purposes to data processing services. Only event responsible, staff, and superusers have the right to reduce and analyse each event-based observational data. User authorisation alone is not enough to perform reduction and analysis on the OP. In line with the capabilities of the OP, the observational data to be uploaded must be in $fits$ format. However, most of the observers prefer to upload their files as one single compressed file such as .zip, .rar, .gz, .7z to make the upload faster. In this case, the authorised user must first decompress the archive. On the OP, there is a function called $extract\_arch$ that uses the $pyunpack$\footnote{\url{https://github.com/ponty/pyunpack}} package for this operation. Since the cameras used by amateur observers produce data in video formats, usually they upload files such as Audio Video Interleave (AVI), Astro Digital Video Data Format (ADV) \citep{hpavlov2020}, Astro Analogue Video (AAV) \citep{barry2015}, SER\footnote{\url{https://free-astro.org/index.php/SER}}. In this case, a video-to-FITS format conversion is required to analyze video frames. Although such a conversion module has been developed, it has not yet been implemented to the OP as a service which is still in the testing phase. The developed module named {\it video2fits} can easily read each video frame and convert it to FITS format, has been developed taking advantage of $stacker.py$ in {\it PyMovie} \citep{anderson2019}.  Moreover, reading the timestamps from the video formats such as AVI, where the observational timestamp is inserted into the video stream directly, is recognised by \textit{easyOCR}\footnote{\url{https://github.com/JaidedAI/EasyOCR}} module, not \textit{PyMovie}'s OCR (Optical Character Recognition) module.

Once all the data is uploaded in FITS format or converted properly to FITS format, OP activates the \textit{Analyze tab} for authorised users, and all data are available for examining via \textit{JS9} tool \citep{eric_mandel_2022_5815246}. Before analysing, the user should first select a best-seeing reference image \citep{alard1998,huckvale2014} amongst the listed frames and locate the target, comparison, and guide (tracking) stars on this reference image.  However, it may be challenging to find the target star in a crowded area. In this case, the OP has a \textit{Locate Occ. Star} button to make the astrometric calibration of the field via \textit{astrometry.net}, which compares quadruple or quintuple geometric hash sets created from detected stars on the image with pre-indexed patterns \citep{lang2010}. The {\it OP} uses the index files for those pre-indexed patterns generated by \cite{kilic2018} from {\it Gaia} DR1 catalogue which is more than adequate to automatically find the target star on the reference image. Moreover the OP will perform the pre-reduction process as long as the pre-reduction option is active from the web interface since images have \textit{IMAGETYP} keywords in the FITS header as Bias, Dark or Flat. Lastly, it is essential to determine the suitable measuring aperture parameters of the target, comparison, and guide stars for the best photometric results.

The images must also be well-aligned to match the target, comparison, and guide stars over entire image series for a successful photometry. On OP, there are different alignment methods in use. Images are aligned with alipy\footnote{\url{https://github.com/japs/alipy}} on the first attempt. If this process fails for any reason, the Astroalign package \citep{beroiz2020100384} is used. If the Astroalign package does not succeed, OP finally triggers another alignment method named $find\_source\_in\_roi$. This method is based on finding the brightest star in the desired \textit{ROIBOX (Region of Interest Box)} which limits the area shifting over box size, indicating the maximum shifting limit over sub-frames. If all alignment attempts fail, the OP system has the ability to do photometry as long as a coordinate file containing three stars' physical coordinates in all images is given. As an example, PyMovie's result file may be used to provide those coordinates. After identifying the sources on each frame, the OP proceeds to the photometry stage. Here, photometry is performed using IRAF's phot task \citep{iraf1986,irafstetson1987} via $myraflib$ \citep{kilic2016}. The user defines the photometric aperture and sky annulus, in the web interface, which are kept fixed through out the process.
Once the photometry process is completed, the photometric results can be displayed for the target (occulted star), reference and guide stars as three light-curves in the Results section and the user is notified by the \textit{Notification Service} via e-mail. These three light-curves displayed using Highcharts\footnote{\url{https://www.highcharts.com/}} are created with the help of the following equations.

Eq. \ref{eq:ft}, where $F_{T( i )}$ denotes the normalised flux of the target star in each frame, $f_{{T}( i )}$ the instrumental flux of the target star measured from the $i^{th}$ frame, and $\overline{f_{T}}$ is the median flux of the target star calculated from all frames.

\begin{equation}
\label{eq:ft}
F_{T( i )}=\frac{f_{T( i )}}{\overline{f_{T}}}
\end{equation}

Eq. \ref{eq:fr}, where $F_{R( i )}$ is the normalised flux of the reference star in each frame $(i)$, $f_{{R}( i )}$ the instrumental flux of the reference star measured from the $i^{th}$ frame, and $\overline{f_{R}}$ denotes the median flux of the reference star calculated using all frames.

\begin{equation}
\label{eq:fr}
F_{R( i )}=\frac{f_{R( i )}}{\overline{f_{R}}} 
\end{equation}

\begin{figure}
    \centering
    \begin{multicols}{3}
        \stackunder{\includegraphics[width=1.3\linewidth]{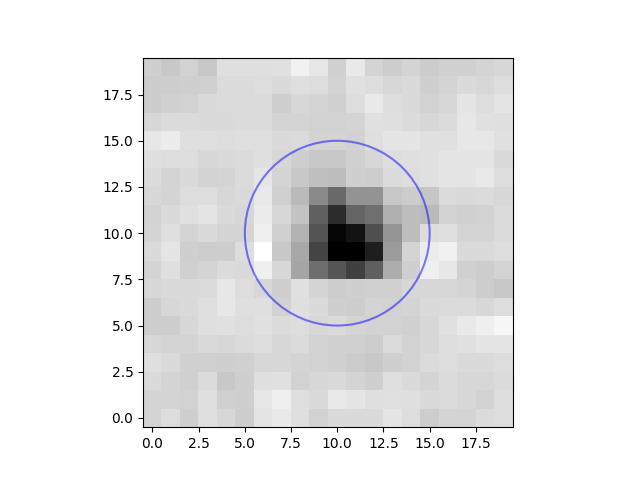}}{Occultation target\label{fig:occ_roi}}\par 
        \stackunder{\includegraphics[width=1.3\linewidth]{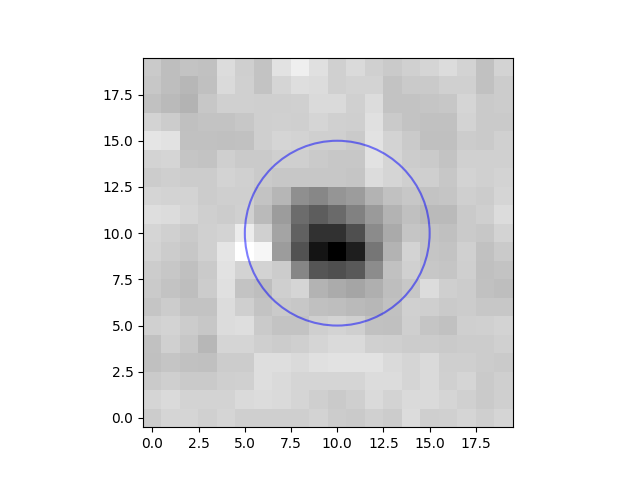}}{Reference star\label{fig:ref1_roi}}\par 
        \stackunder{\includegraphics[width=1.3\linewidth]{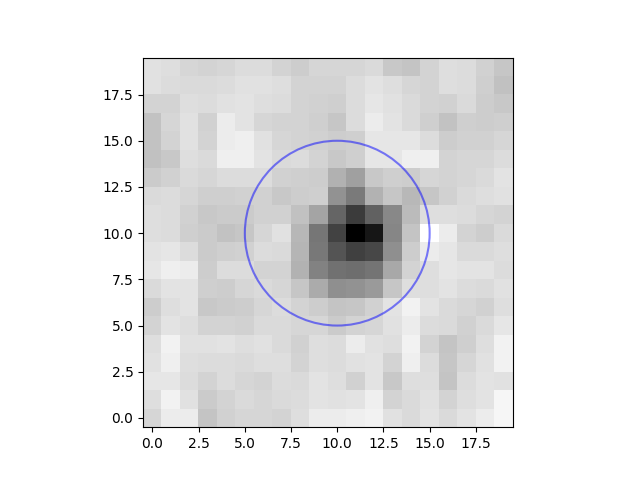}}{Guide star\label{fig:guide_roi}}\par 
    \end{multicols}
    (a)
    \begin{multicols}{3}
        \stackunder{\includegraphics[width=1.3\linewidth]{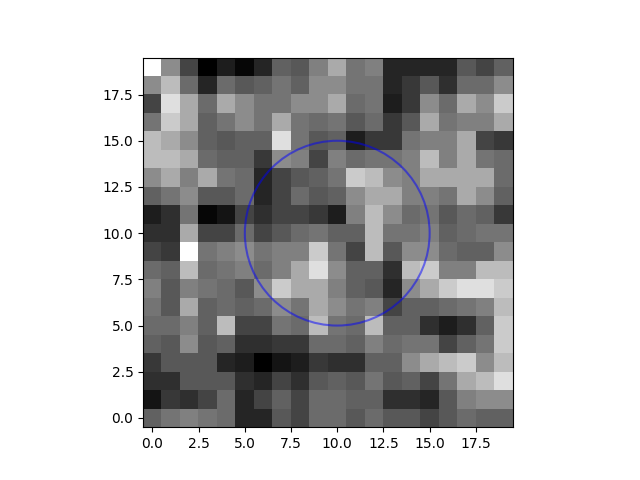}}{Occultation target\label{fig:occ_roi_event}}\par
        \stackunder{\includegraphics[width=1.3\linewidth]{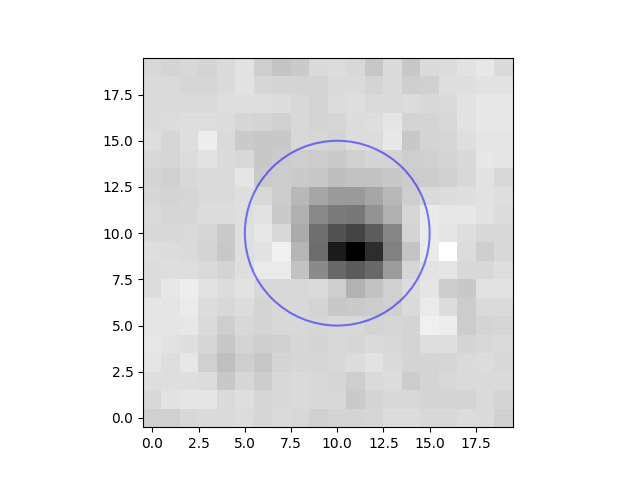}}{Reference star\label{fig:ref1_roi_event}}\par
        \stackunder{\includegraphics[width=1.3\linewidth]{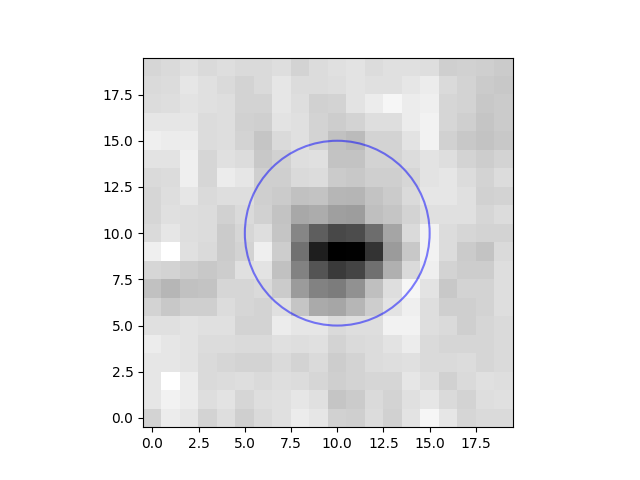}}{Guide star\label{fig:guide_roi_event}}\par 
    \end{multicols}
    (b)
\caption{This figure shows the target, reference and guide stars detected by the OP within 20 pixels of \textit{ROIBOX} in each frame. In Figure \ref{fig:roiboxes}a, the state of three stars is shown in a series of images where the occultation event does not happen. In Figure \ref{fig:roiboxes}b, the state of 3 stars during the occultation moment is shown. As can be seen, the occultation target is not detected in Figure \ref{fig:roiboxes}b as the target star was occulted by the solar system object, the former being too faint to be detected in the exposure.}
\label{fig:roiboxes}
\end{figure}

Eq. \ref{eq:fn}, where $F_{N( i )}$ represents the ratio of $F_{T( i )}$ and $F_{R( i )}$ in each frame. This ratio mainly eliminates the variations on the target star due to atmospheric and instrumental effects which cause a false variation in flux measurement. Nonetheless, as shown from Fig.\ref{fig:roiboxes}, the OP allows users to visualise the \textit{ROIBOX} where the photometry is applied to check whether it is an occultation event itself or not by clicking any point on the light-curve graph (see appendix \ref{apd:op_light_curves}).

\begin{equation}
\label{eq:fn}
F_{N( i )}=\frac{F_{T( i )}}{F_{R( i )}} 
\end{equation}

Finally, the user is now able to export the light curve (LC) in various formats such as graphics or ASCII files. The ASCII LC file is specially designed to be compatible with the SORA (Stellar Occultation Reduction and Analysis) software \citep{altair2022} to further exploit the data and characterise of the occulting object.


\subsection{The Notification Service}\label{sec:notification_service}


The Notification Service informs the user and/or superuser(s) at the start or end of most running processes on the OP. All notifications such as user management (registration, activation, account update, etc.), archive data requests, data analysis processes, announcements, user issue tracking, and server notifications are carried out by the Notification Service over the SMTP (Simple Mail Transfer Protocol) server via transactional e-mails.

\subsection{Backup service}\label{sec:backup_service}

The database on the application server seen in Fig. \ref{fig:occultation_platform_design} is backed up to the NAS (Network-attached storage) server with the django-dbbackup tool\footnote{\url{https://github.com/jazzband/django-dbbackup}}. After that, the data on the NAS is cloned to an asymmetric backup server daily with the rsync tool\footnote{\url{https://rsync.samba.org/}} incrementally. 


\section{Results}\label{sec:results}

\begin{figure}
\centering
\includegraphics[width=\columnwidth]{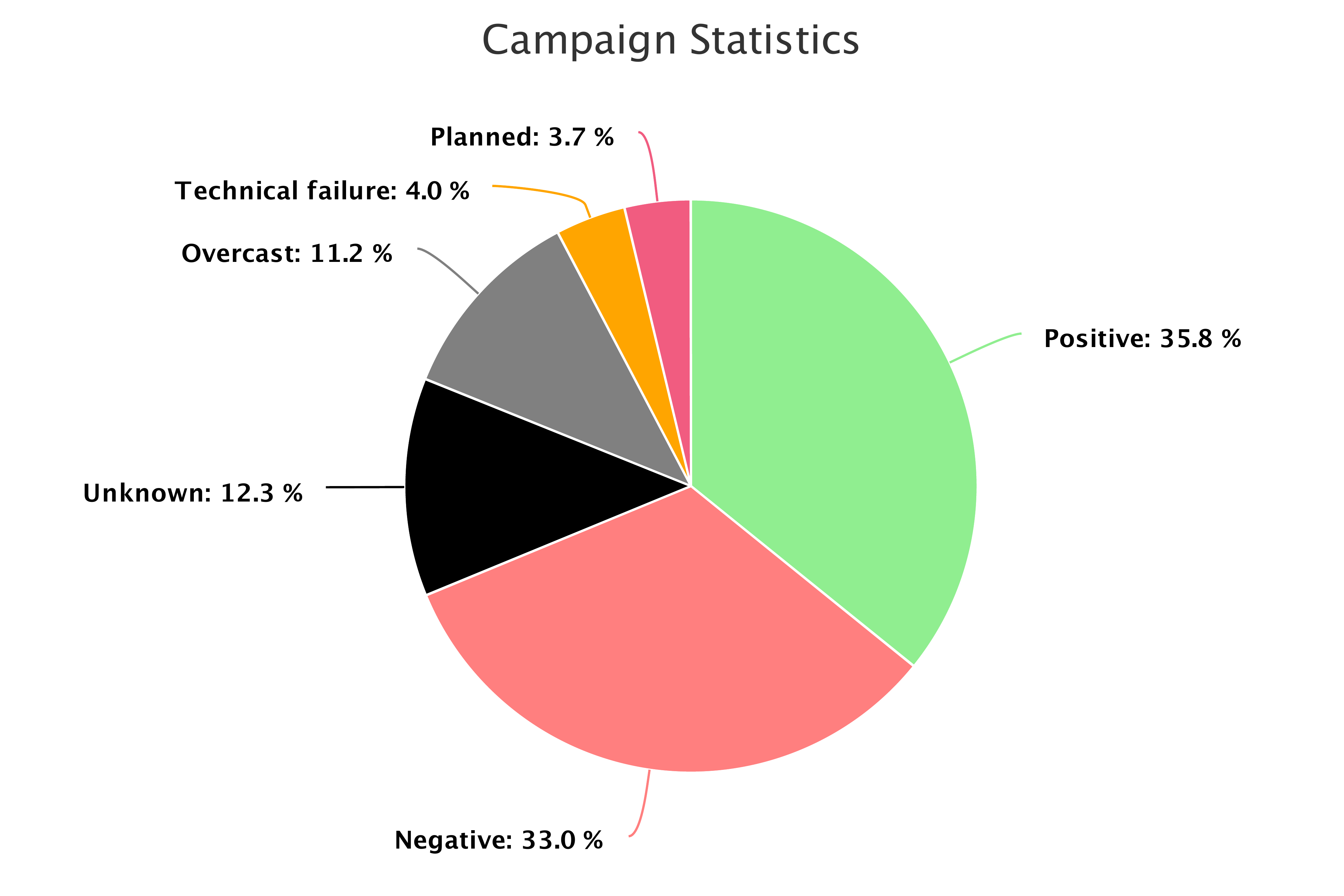}
\caption{General statistics of campaigns run between Aug. 2020 and May 2022. In the chart, \textit{Positive} means that at least one data set has disappearance of the target star because of occultation event while \textit{Negative} means the none of the recorded data shows the disappearance of the star. \textit{Planned} means an observer did not provide any information about the event despite being planned to observe. \textit{Technical failure} means that the observer experienced a technical issue at the moment of the occultation. \textit{Overcast} means observation could not be conducted due to bad weather conditions. \textit{Unknown} means there was no information, although data were provided.}
\label{fig:campaign_statistics}
\end{figure}

The Occultation Portal (OP) has officially been in service since the 2002 MS$_4$ TNO occultation campaign event occurred on 8 August 2020, but as it was a test case, it was used by a restricted number of observers. As of 23 May 2022, 249 different occultation predictions provided by the Lucky Star Project and its collaborators have been announced on OP as occultation campaign calls. For 115 out of those 249 campaigns, data is provided by at least one observer (chord). Due to various reasons such as bad weather conditions, inaccurate predictions, observations from outside the occultation path, insufficient star brightness etc., it is not expected to see the occultation event from all these data. 35.8\% of the events were reported as Positive, which means the disappearance of the target star can be seen because of the occultation event. All these statistics are shown in detail in Fig. \ref{fig:campaign_statistics} which can be accessed at any time via the OP\footnote{\url{https://occultation.tug.tubitak.gov.tr/statistics/}}.

\begin{figure}
\centering
\includegraphics[width=\columnwidth]{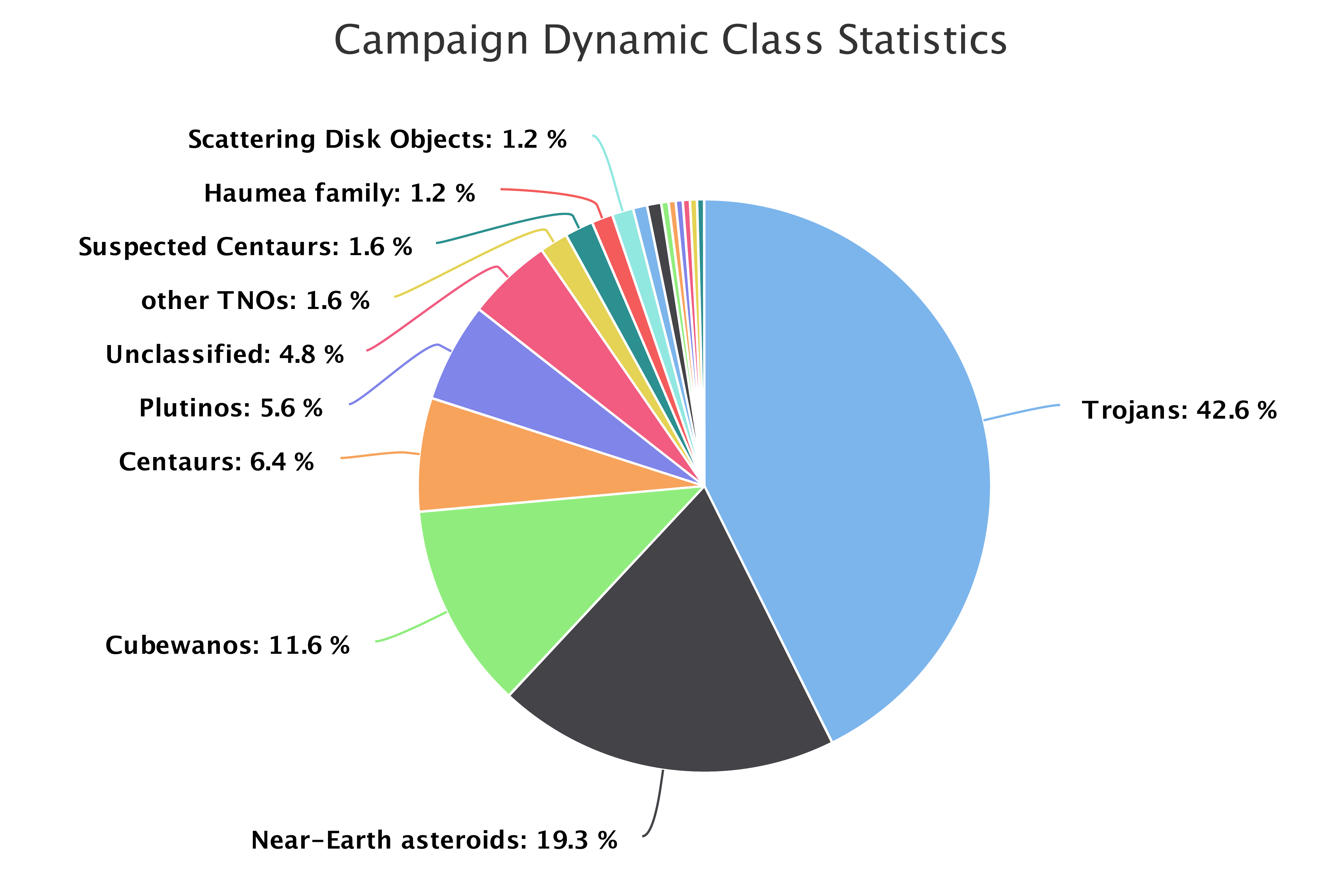}
\caption{Statistics on the dynamical classes of the occulting objects that were announced predictions on OP between Aug. 2020 and May 2022. For the sake of clarity, the names of the dynamical classes that contribute to less than 1\% are not shown in the chart.}
\label{fig:campaign_dynamic_classes}
\end{figure}

Another important statistic that should be given is that of the dynamical classes of the objects that have been announced on the OP between Aug. 2020 and May 2022. It is clearly seen from Fig.\ref{fig:campaign_dynamic_classes} that the vast majority of the objects are beyond the main belt, which is mainly because the Lucky Star project focuses on primordial bodies. These objects are the unaltered remnants of the protoplanetary disc that existed in the early stages of our Solar System, they can thus supply us with critical information and invaluable constraints about the evolution of our Solar System \citep{lykawka2008,morbidelli2008}. In Fig.\ref{fig:campaign_dynamic_classes} NEAs are also seen at a proportion of $\sim$20\% since the inclusion of occultation predictions of the ACROSS project\footnote{\url{https://lagrange.oca.eu/fr/goals}}, whose primary goal is to improve the position accuracy of the NEAs in the Gaia frame.

\begin{figure*}
\centering
\includegraphics[width=1.0\textwidth]{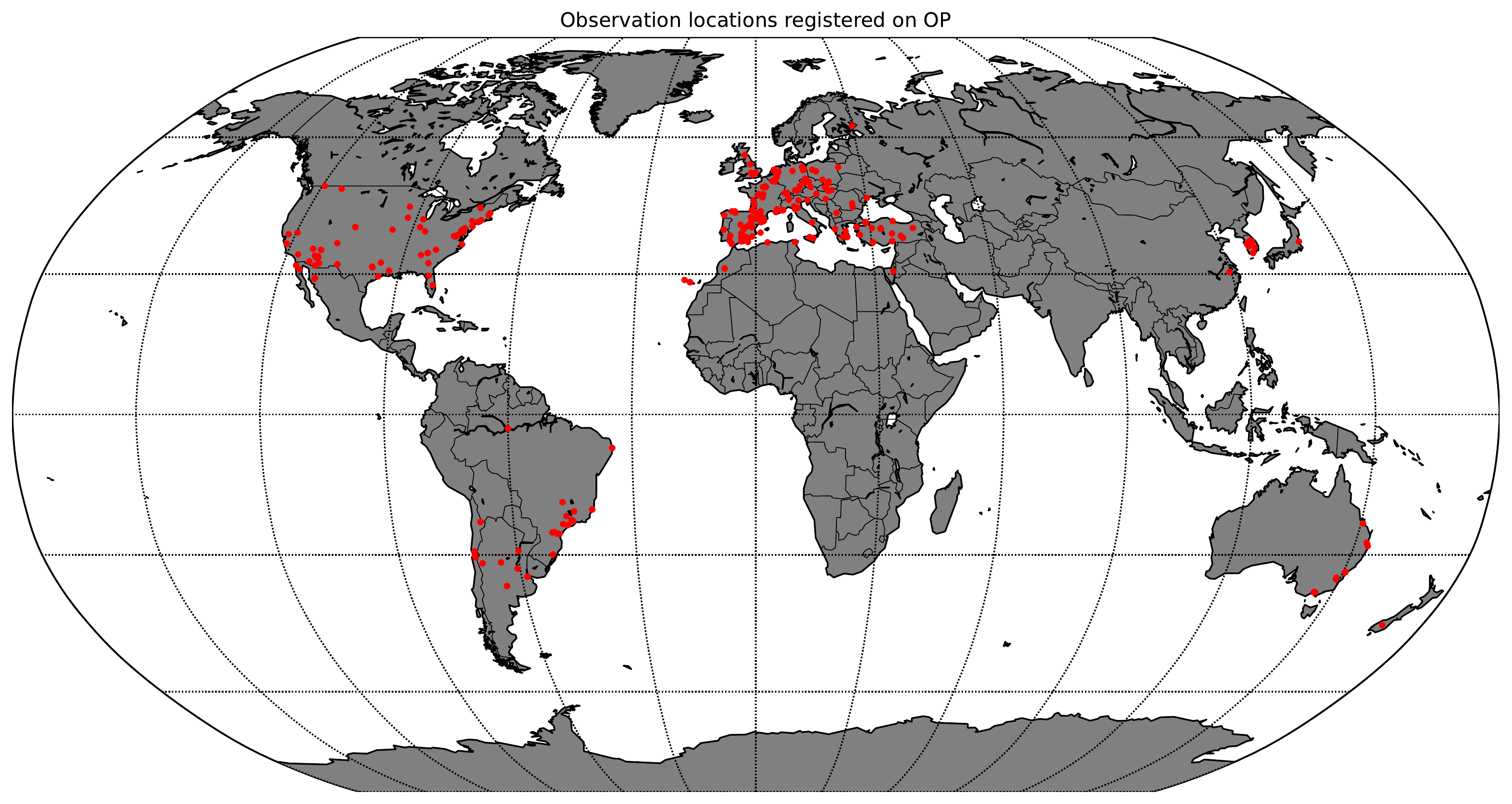}
\caption{Occultation Portal: observers network. A total of 331 red dots indicate the observers' locations registered on the OP. These could show either mobile stations or fixed telescopes. It is clearly seen that  most of the locations are in Europe, while others are spread over North Africa, North America, South America, far East Asia, some parts of Australia. As occultation events are announced in different parts of the world, we expect that the number of registered observers to increase in the future and red dots to spread out.}
\label{fig:occultation_portal_locations}
\end{figure*}

\begin{figure}
\centering
\includegraphics[width=\columnwidth]{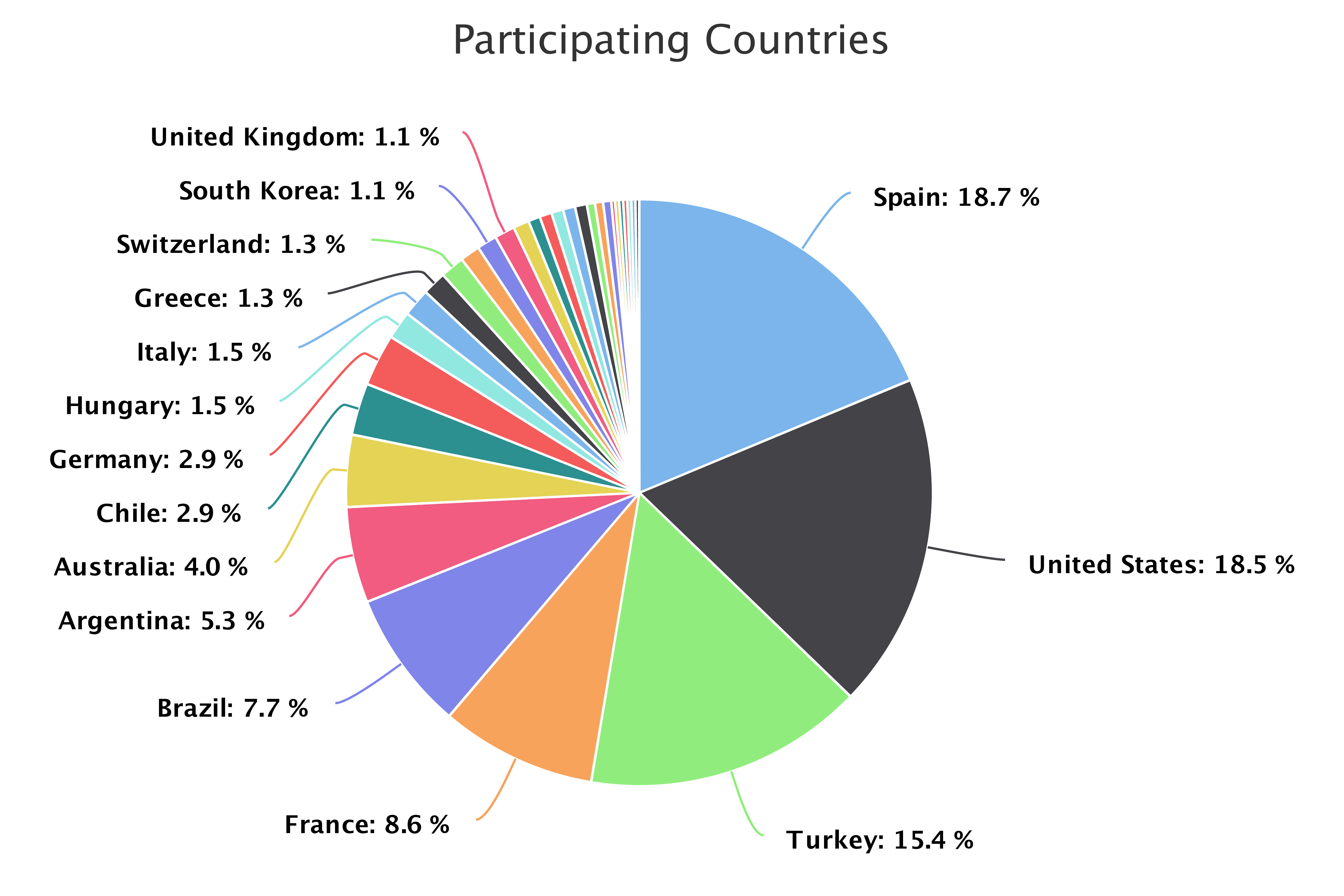}
\caption{Occultation Portal participating countries so far between Aug. 2020 and May 2022. For the sake of clarity of the figure, the names of the countries contributing less than 1\% are not shown in the chart.}
\label{fig:occultation_portal_locations_by_countries}
\end{figure}

The campaign conducted on OP with the highest participation of 56 chords was the occultation by TNO 2002 TC302 (84522) on 11 Nov 2021\footnote{\url{https://lesia.obspm.fr/lucky-star/occ.php?p=75799}}. Together with this successfully conducted campaign, it can be expressed that OP has become an essential platform for occultation research. As of 23 May 2022, there are 301 registered users on the OP. The number of observation locations recorded by these users is 331. These locations are shown on the map in Fig. \ref{fig:occultation_portal_locations}. Besides, the status of these observers based in their countries is given in Fig. \ref{fig:occultation_portal_locations_by_countries}, it should be noted that these graphs may change over time.

\begin{figure}
\centering
\includegraphics[width=\columnwidth]{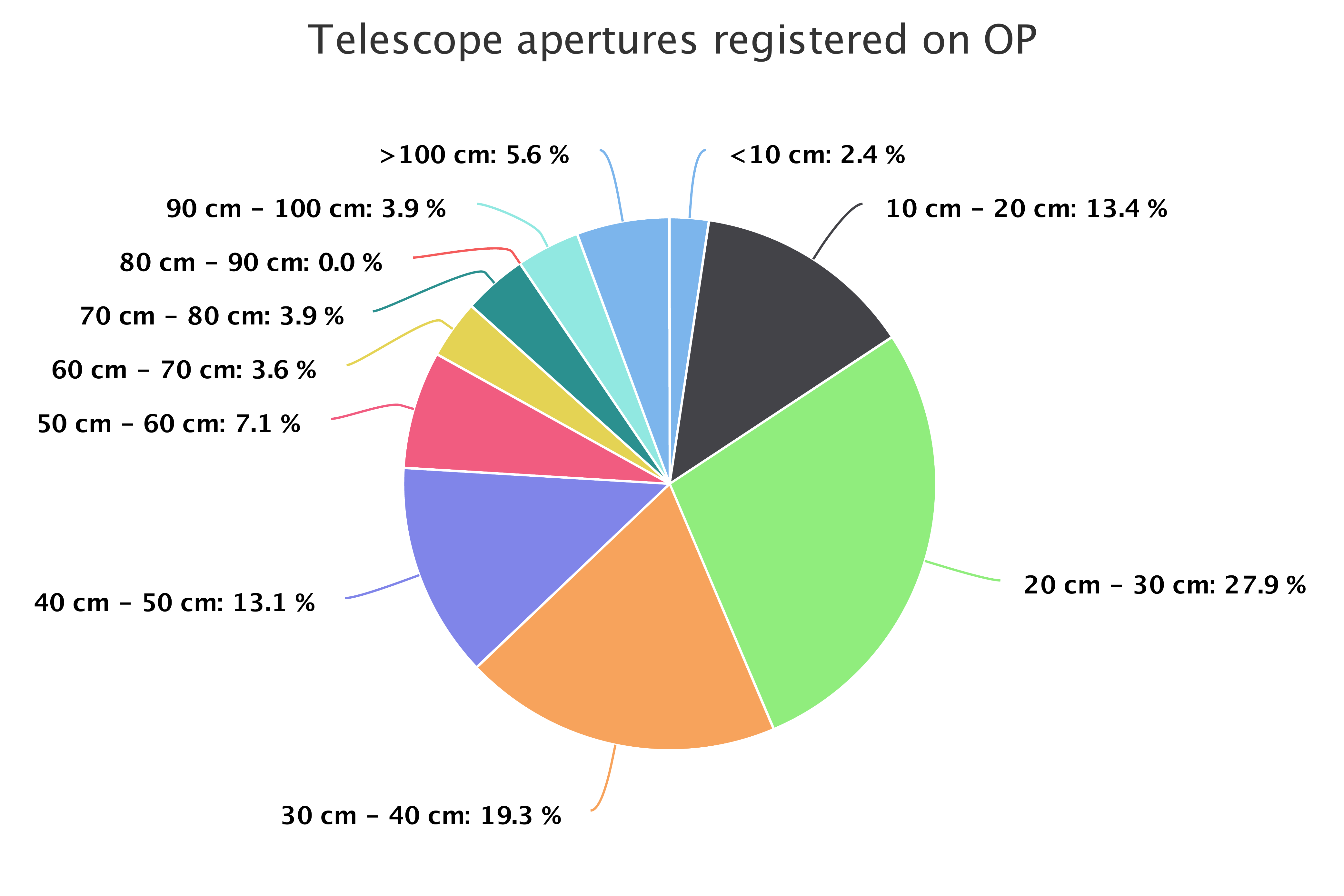}
\caption{Telescope diameters registered on OP between Aug. 2020 and May 2022. In the chart, the most notable point is that most of the registered telescope diameter (> \%75) are less than 50 cm in diameter. It means that stellar occultation observations can be considered independent of telescope diameter as long as the occulted star is bright enough.}
\label{fig:occultation_portal_telescopes}
\end{figure}

Another interesting result from the OP is the number of telescopes by aperture size used in the campaigns. Statistics of the number of telescopes by aperture size generated from the information provided by those 301 registered users are represented in Fig \ref{fig:occultation_portal_telescopes}. These statistics show us that, the biggest proportion ($\sim$28\% of the registered telescopes are small telescopes with diameters in the range of 20-30 cm. Furthermore, most of the registered telescope apertures are less than 50 cm ($>75\%$). It can be clearly stated that as long as the occulted star is bright enough\footnote{\url{https://occultation.tug.tubitak.gov.tr/statistics/}}, relatively small aperture telescopes are of paramount importance for stellar occultation observations. This is also an indication that the substantial number of participants taking part in the campaigns are amateur astronomers. 

\begin{figure}
\centering
\includegraphics[width=\columnwidth]{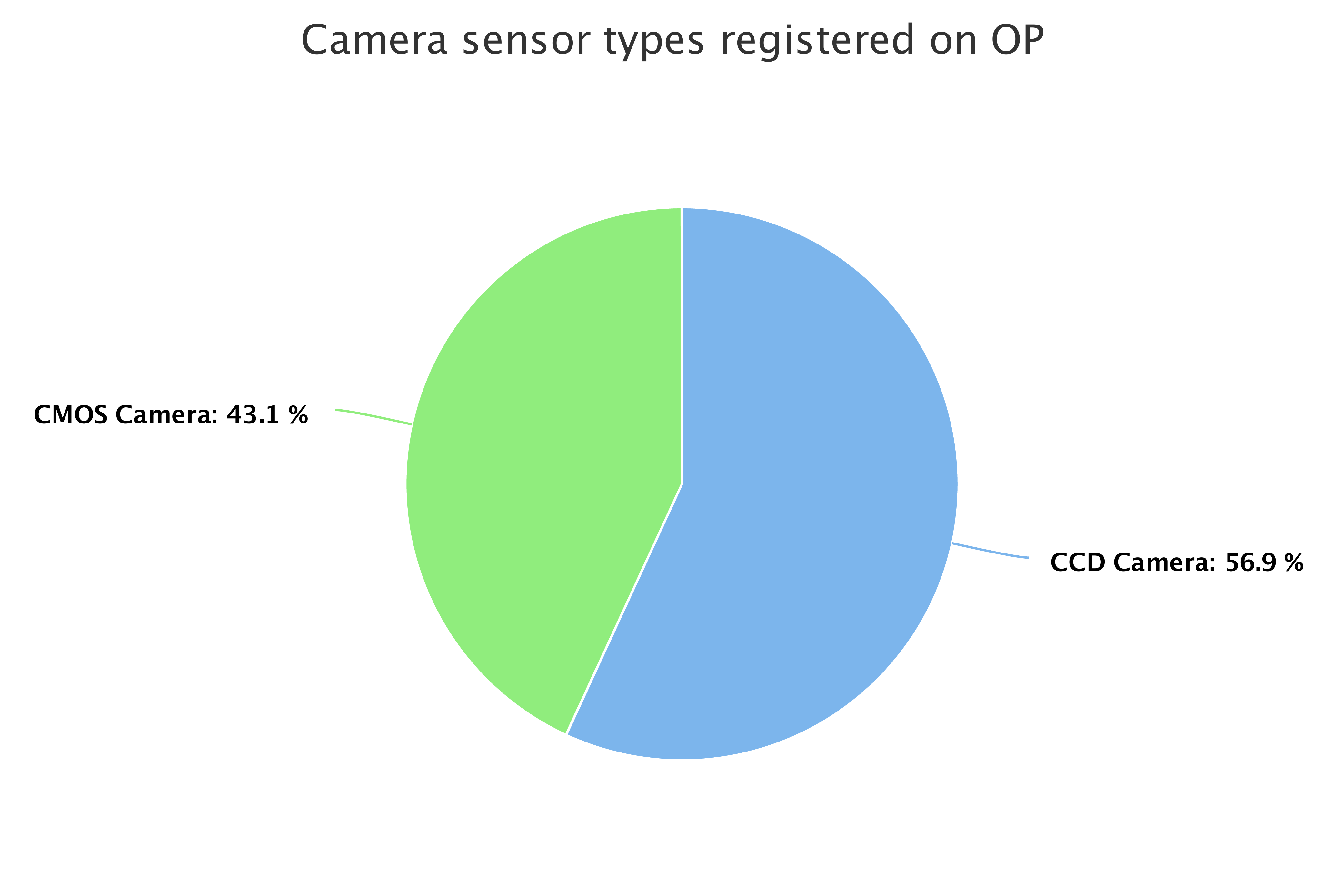}
\caption{Camera types registered on OP between Aug. 2020 and May 2022. CCD Camera refers to an astronomical detector with a CCD sensor produced mainly by Andor, FLI, Apogee, SBIG and, Watec, etc. while CMOS Camera refers camera that uses a CMOS-based image sensor most observers are using, such as QHY cameras, RunCam, ZWO ASI, etc.}
\label{fig:occultation_portal_cameras}
\end{figure}

In Fig.\ref{fig:occultation_portal_cameras}, it is seen that 56.9\% of the cameras attached to the telescopes are CCD (Charge-coupled device) cameras and $\sim$20\% of them consist of only Watec video cameras, followed by CMOS (Complementary metal-oxide-semiconductor) cameras (43.1\%). It can be seen that the number of CCD and CMOS cameras are close to each other. This shows that not only CCD cameras but also CMOS cameras are now contributing on observational studies.

\begin{figure*}
\centering
\includegraphics[width=\textwidth]{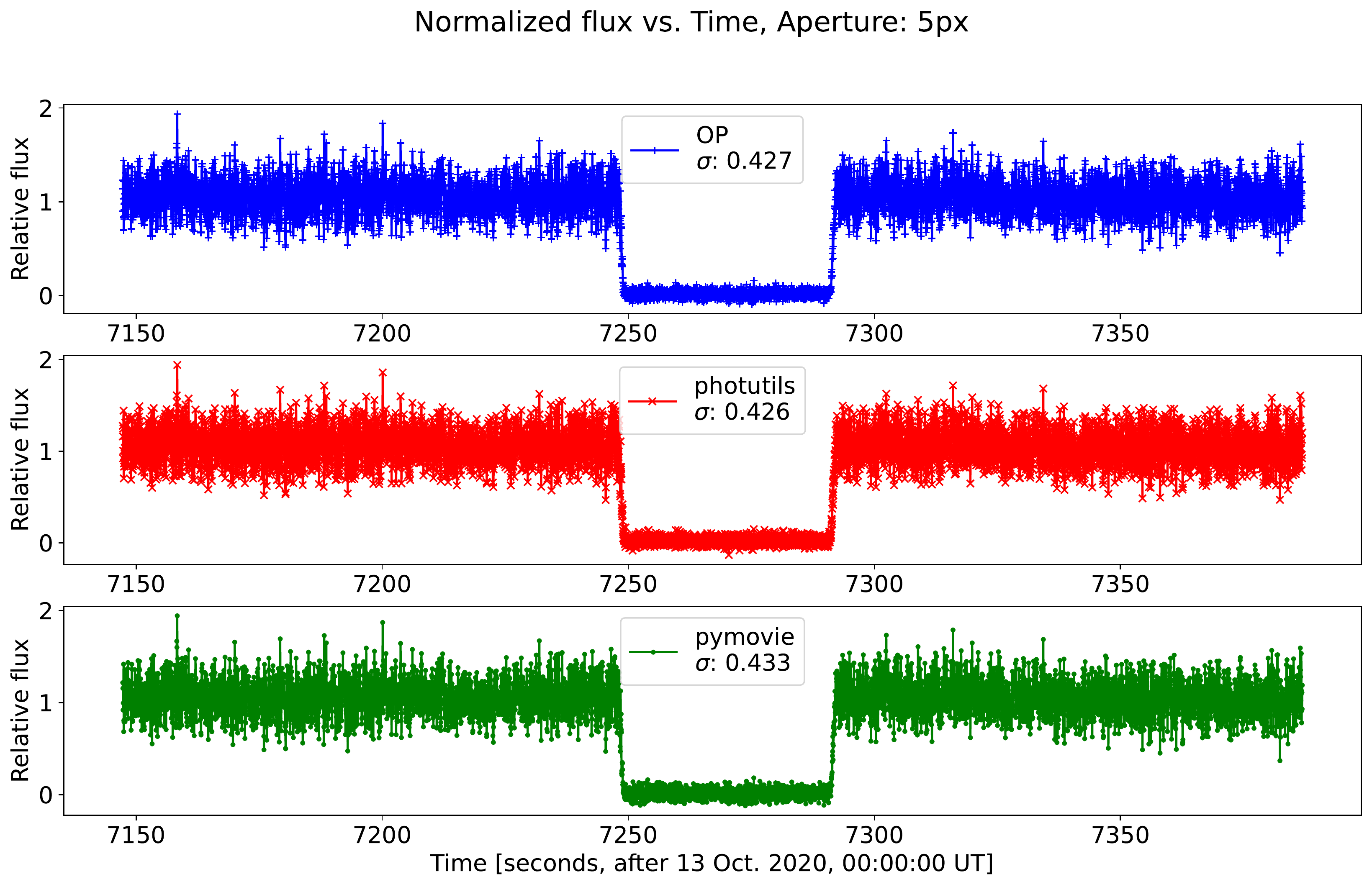}
\caption{Data analysis and reduction service benchmarking. This graphic was created as a result of photometry of 7184 frames. It shows the light curves for a positive stellar occultation observation of a large TNO 28978 Ixion recorded by Ted Blank on 13 October 2020. The light curve with the blue plus markers at the top is obtained by using {\it OP}'s photometric reduction method. The red one in the middle with cross markers was obtained with \textit{photutils} while the green one with dot markers at the bottom was obtained with \textit{PyMovie}. The $\sigma$ represents the standard deviation in the light curve in the graphs.}
\label{fig:photometry_tools_comps}
\end{figure*}

One of the most important services of the OP is {\it Data analysis and reduction service} where photometric results are produced. To measure the reliability of the photometric results \textit{photutils} \citep{larry_bradley_2020_4044744} and \textit{PyMovie} were used.The results were compared, using data from the observation of the occultation by Ixion on October 13, 2020\footnote{\url{https://lesia.obspm.fr/lucky-star/occ.php?p=41312}}, and the results obtained with all tools are plotted in Fig. \ref{fig:photometry_tools_comps}. As can be seen in Fig. \ref{fig:photometry_tools_comps}, the standard deviation ($\sigma$) of the photometric analysis results are not much different than the others. 

Finally, the Occultation Portal runs on Pardus (19.3), a Debian-based GNU/Linux operating system developed by the Turkish Academic Network and Information Center (ULAKB\.{I}M). It was in service when the 2002 MS$_4$ campaign event took place on August 8, 2020. Since the launch of OP, the MTD (Mean Tolerable Downtime) of the system is less than one hour. The system has been down six times since being in action, however, that was all due to scheduled maintenance.

\section{Discussion and Conclusions}\label{sec:conclusions}

In this paper, we presented the details of the OP's current system architecture and workflow in detail. The current version of the OP (v2.11-beta) has all the components to support the occultation science on campaign announcement, data collection, archiving, management and reduction to produce results efficiently. With hundreds of events, it has been demonstrated that the OP efficiently stores observational information while easing data archiving, analysis, and management. On the long run, observational data and database size on OP will increase considerably, and improved data science methods will have to be used to store and process this data effectively and efficiently. Considering the extensive impact of the {\it LSST} and {\it Gaia}'s subsequent releases on occultation observations, it might also be an excellent solution to make data of the OP public later following IVOA standards. \citep{ivoa2010}.

Primarily the OP was a customised platform only for the Lucky Star campaigns; but now, with the implementations made after the participation of the ACROSS project to the OP, any prediction with the {\it occultation elements file} in the XML format on the OW Cloud can be added/announced on the OP by the authorised users. Yet, the OP can become a common stellar occultation data collection centre of all groups, such as those mentioned in Section \ref{sec:intro}. Along with a common central server, another plan is to make the source code available under a free software licence (e.g. GNU GPL v3) for the stand-alone server deployments of the OP for other research organisations and teams.

Being the OP served from a single location, the time to upload large-scale observational data will increase for the users on far locations, such as different continents. So, to prevent this and limit the time loss, it has been planned to deploy the on-premise mirror servers where located in different zones. Besides, the system will have better high-availability and higher fault-tolerance. The other option to achieve these goals is to migrate all systems to a Cloud Platform. These plans heavily depend on the continuous funding of the platform, budget and labour force. OP only has the Prediction service API, but in the future, it is planned to offer all services as API or with a client running on the terminal.

In stellar occultation campaigns, it is obvious that small telescopes make great scientific contributions. Accordingly, OP is a great platform to gather these small telescopes around the world, which can together cause an extraordinary impact on citizen science. As a result, with this system, we especially encourage amateur astronomers to use their telescopes for scientific purposes, and thus we offer a non-profit, scientific and time-saving practical portal that effectively brings together both researchers and amateur astronomers.

\section*{Acknowledgements}

We would like to thank T\"UB\.{I}TAK National Observatory (Turkey), Akdeniz University (Turkey), The Federal University of Technology – Paraná (Brasil), and Lucky Star Project/Paris Observatory/LESIA (France) for their support of this project. This work was carried out within the Lucky Star umbrella that agglomerates the efforts of the Meudon, Granada and Rio teams, which is a European Research Council project under the European Community's H2020
2014-2021 ERC Grant Agreement no. 669416. This study was financed in part by the Coordenação de Aperfeiçoamento de Pessoal de Nível Superior - Brasil (CAPES) - Finance Code 001. FBR acknowledges CNPq grant 314772/2020-0. BEM acknowledge the CNPq grant 150612/2020-6. Additionally, we would like to thank the member of the International Occultation Timing Association (IOTA), Ted Blank (US), for providing observational data for the study. Thanks to H. Aziz Kay{\i}han for proofreading the article. The authors thank the referee Hristo Pavlov for the constructive comments and recommendations that helped us improve the manuscript. Pie charts used throughout the article was created by Highcharts.

\section*{Data Availability}\label{sec:data_availability}

All assets, data, and documents used in this article are available on \url{https://occultation.tug.tubitak.gov.tr/} and \url{http://occultationportal.org/}. Source code may be shared on reasonable request to the corresponding author.



\bibliographystyle{mnras}
\bibliography{references} 




\appendix

\section{Supplementary figures}\label{sec:supplementary_figures}
Some screenshots of the {\it OP} front-end (v2.02-beta) are provided for illustrative purposes to interest the reader.

\begin{figure*}
\centering
\includegraphics[width=\textwidth]{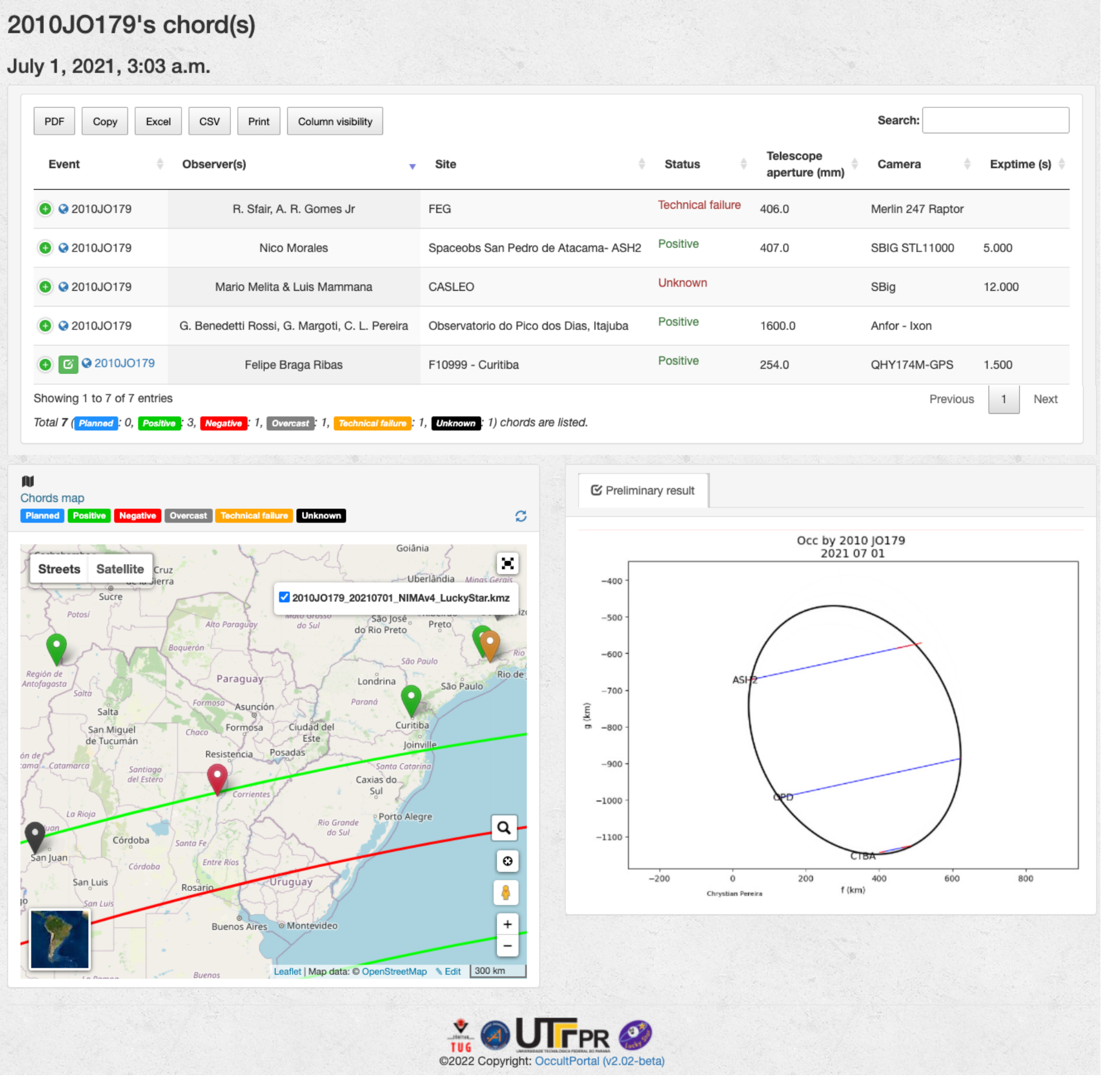}
\caption{The front-end of the chords report page. The successful occultation campaign by the TNO (574372) 2010 JO$_{179}$ dated July 1, 2021, is chosen as an example. Each observer can see the information of each reported observation, and can edit his own information previously provided. The lower-left panel displays a map showing the positions of the observers across the predicted shadow path with colors representing the reported result. The lower-right panel presents an image which can be uploaded by the {\it staff} or {\it event responsible}. Usually it presents a preliminary limb fit obtained from that campaign.}
\label{apd:op_chords_report_page}
\end{figure*}

\begin{figure*}
\centering
\includegraphics[width=\textwidth]{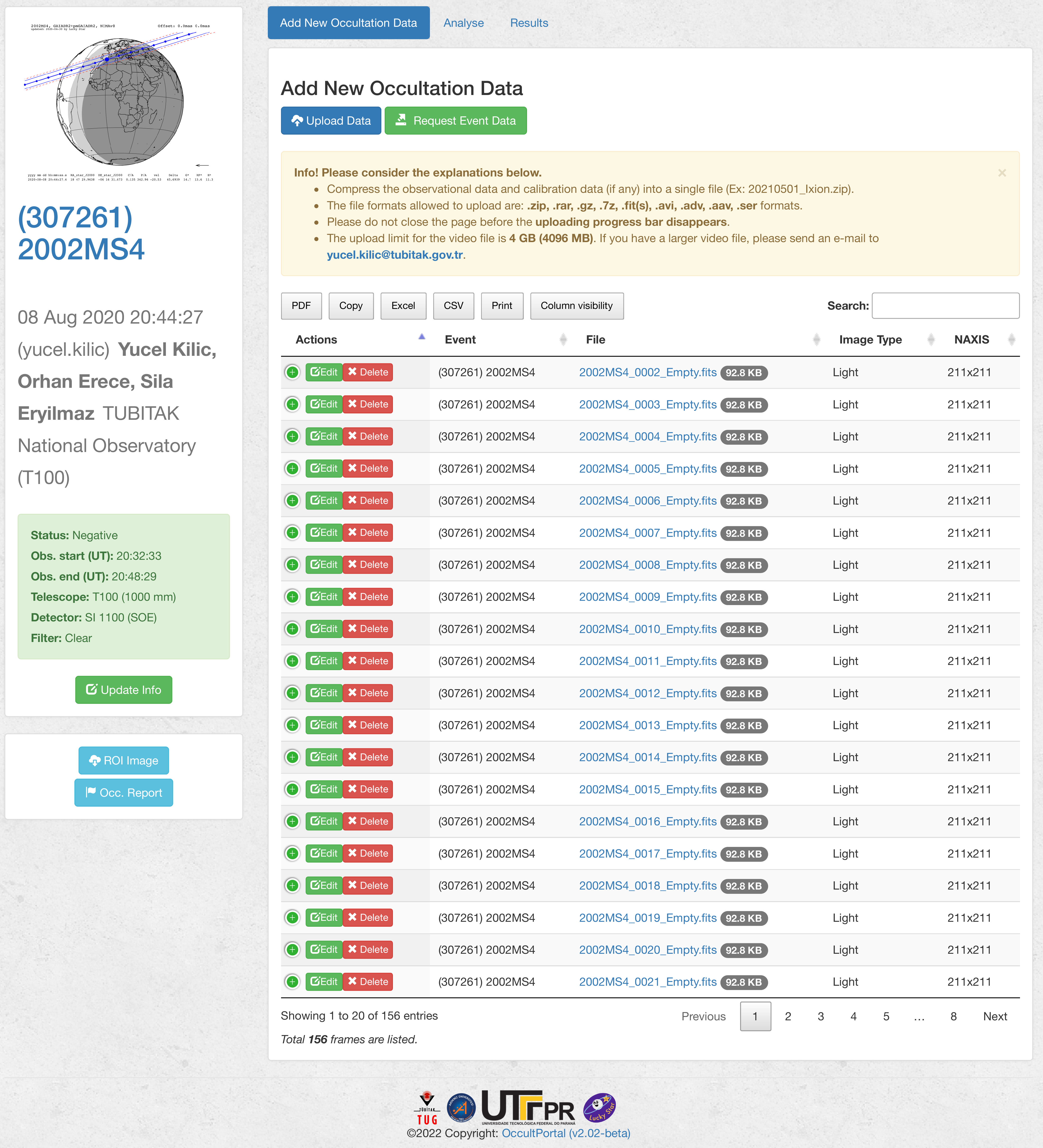}
\caption{A sample report page of an observer who provided data to the observation campaign for the occultation by Ixion on August 8, 2020. On this page, the observer and all authorized users can upload data with the {\it Upload Data} button, request all reported information and data with the {\it Request Event Data} button, update information about the observation with the {\it Update Info} button, view the additional files uploaded by the observer with the {\it Additional Files} button, and view the status report of the observation in Asteroidal Occultation Report format with the {\it Occ. Report} button.}
\label{apd:reported_observation}
\end{figure*}

\begin{figure*}
\centering
\includegraphics[width=\textwidth]{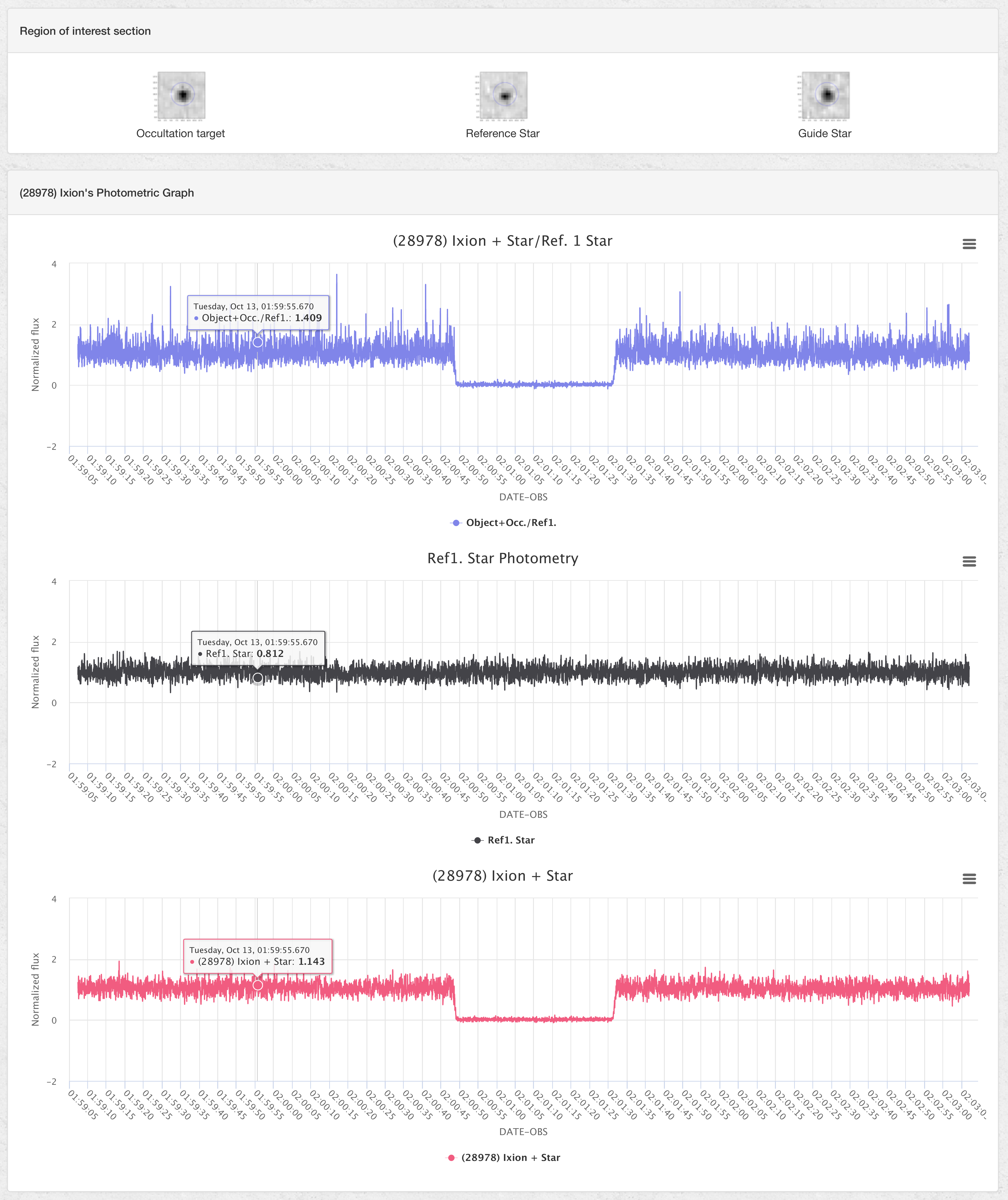}
\caption{Aperture photometry result page displaying the obtained Light Curves. The small images at the top of the figure change according to each clicked point on the graphic. Thus, the user can visually examine the occultation event. The blue graph shows the light curve ratio of the "target star + Solar System object" to the reference star, the black graph shows the normalised light curve of the reference stars, and the red graph shows the normalised light curve of the "target star + Solar System object" only.}
\label{apd:op_light_curves}
\end{figure*}



\bsp	
\label{lastpage}
\end{document}